# ELECTROSTATIC FIELD OF CHARGED INFINITE METALLIC PLANE CONTACTING WITH METALLIC SPHERE

P. M. Mednis

Novosibirsk State Pedagogical University, Chair of the General and Theoretical Physics, Russia, 630126, Novosibirsk,Viljujsky, 28
e-mail: pmednis@inbox.ru

**Abstract**

The problem concerning the form of the electrostatic field potential of the infinite charged metallic plane contacting with the metallic sphere is solved. The distributions of the surface charge densities both on the plane and the sphere and also the force acting on the sphere that causing to tear away from the plane are found. The expressions found for the charge of the sphere and the force of tear from the plane permits estimate the frequency of the shuttle motion of the nanoparticles and the charge transfer between metallic electrodes and also use them for the measurement of their sizes and to make use of their separation in progress.

## 1. Introduction

The classical methods of solving the electrostatic problems are widely known [1a,1b]. As usual they are illustrated by means of a lot of practically important examples and tasks [1a,1b,2]. There is no among them the explicitly context problem of defying an electrostatic field of the infinite charged metallic plane contacting with the metallic sphere. Nowadays in course of a nanotechnology development the solving of this problem turns out to be actual.

Indeed, in the practical sense because of the pushing off forces the effect of tear away the sphere from the plane takes place. For the macroscopic sphere rebound from the plane in case of it's vertical orientation and the corresponding charge transfer to opposite charged contact has been described as early as in [3]. The effect of the jumping spheres with diameters near 0.5 - 1 mm in case of horizontal orientation of the plane is widely known as an object of standard lections demonstrations in the electromagnetism course. In technical and scientific sense this jump effect indirectly connected with the work [4] in which the defined mechanism of the directed charge transfer between the two electrodes was suggested. In particular the aim was of creating a new type of the transistor. As follows from the report in [5] this new effect was realized experimentally recently in [6]. There were used gold nanoparticles with diameters 20 nm contacting with gold nanoelectrodes by virtue of non conducting flexible molecules. The shuttle process was initiated after applying a sufficiently large bias voltage to electrodes. Because of the tunnel effect gold nanoparticles were charging from the minus electrodes transferring then the charge to the positive electrode. After loss of a part of the charge nanoparticles returned back and process repeated with the frequency of elastic vibration by the value of order $10^{11}$ Hz.

It is shown in the present work that high frequency charge transfer is possible also because of the jumping effect mentioned above. In addition the results of the problem solved may be applied for some metrological aims slightly discussed below.

In this work the boundary problem under consideration is solved analytically in quadrature form. From the results of computer experiments the distribution of the field potential is found outside the plane and the sphere. It is considered also another characteristics needed of the practical interests.

Note that the most near is the similar problem of two equal contacting metallic spheres solved in [2] by the inverse method. The generalization of this task to the spheres of different radiuses aiming then to limit one of them to infinity is possible in principal. But it is not in fact a trivial task. Because of the standard method of the separation of the variables was used. The solution obeying boundary conditions as a superposition of the separate solutions has been constructed.

## 2. The solution of boundary problem for the field potential

Let us choose the system of references so the *xy* - axes be lie in a conducting plane given. The *z*-axis is perpendicular to this plane and it is passing through the center of the sphere and initial point *O*, which is the contacting point of a sphere and a plane also. Because of the symmetry of the field relative to rotation around the polar axis *Oz*, it is sufficient for us to consider the main Laplacian's equation $\nabla^2 \varphi = 0$ for the potential $\varphi$ of the field only in the arbitrary plane containing the polar axis and which to be perpendicular to the *Oxy* plane. In this case the potential $\varphi$ depends on the distances $r = OP$ to the point of the field and the angle $\theta$ between the polar axis and *OP* line. The angle $\theta$ is changing into interval $0 \leq \theta \leq \pi/2$. So we can put the form $\varphi = \varphi(r, \theta)$. The solution of the equation

$$\sin\theta \frac{\partial}{\partial r}\left(r^2 \frac{\partial \varphi}{\partial r}\right) + \frac{\partial}{\partial \theta}\left(\sin\theta \frac{\partial \varphi}{\partial \theta}\right) = 0 \tag{1}$$

is needed to obey boundary conditions, which are formulated below. Let us find the solution as the sum of the homogeneous part and the additional or disturbing part of the field in the form

$$\varphi = -E_0\, r\cos\theta \cdot (1 - f(\cos\theta) \cdot g(r)) \tag{2}$$

The $E_0$ factor here means the positive electric field produced by the system far from the sphere. The substitution of the expression (2) into equation (1) and separation of the variables with parameters $\lambda$ yields the following equation for the function $f(u)$

$$u(u^2 - 1)f'' + 2(2u^2 - 1)f' + \lambda u f = 0. \tag{3}$$

Here we denoted the $u = \cos\theta$. For the function $g(r)$ we get also the equation

$$r^2 g'' + 4rg' + \lambda g = 0. \tag{4}$$

The general solutions of the equations (3) and (4) are well known [7]. By means of this solution we must construct the additional solution, which is restricted in region $0 < u < 1$. Additionally if the distance $r \to \infty$ this solution must vanish. The analysis get up the such solution realized if one choose $\lambda = -2\,n\,(3 + 2n)$ with $n = 0, 1, 2\ldots$. Then the solution (2) may be written as a sum of the separate solutions

$$\varphi = -E_0 r \cos\theta \left[1 - \sum_{n=0}^{\infty} A_{2n+1} \frac{P_{2n+1}(\cos\theta)}{\cos\theta} \frac{1}{r^{2n+3}}\right]. \tag{5}$$

Here $P_{2n+1}(\cos\theta)$ are Legandr's polynomial with the odd indices. The $A_{2n+1}$ constants must be defined from the boundary conditions. Evidently in all points of the plane which obey the condition $\theta = \pi/2$ the solution (5) vanish. If the points of the observation is lying on the sphere, then $r = l$, where $l = 2R\cos\theta$ is the chord. That is the length of the line from $O$ point to the point of crossing that line with the sphere. The substitution of the $r = 2R\cos\theta$ into equation (5) follows condition that the potential is zero in all points of the surface of the sphere. That is formally takes place the condition

$$\cos^3\theta = \sum_{n=0}^{\infty} \frac{A_{2n+1}}{(2R)^{2n+3}} \frac{P_{2n+1}(\cos\theta)}{\cos^{2n+1}\theta}. \tag{6}$$

The resolution of the left and right sides of the expression (6) into the series by degrees of $\mathrm{tg}^2\theta$ with taking into account the identity [8]

$$P_n(\cos\theta) = \frac{1}{\pi}\int_0^{\pi}(\cos\theta + i\sin\theta\cos x)^n\,dx \tag{7}$$

and the comparison of the coefficients in equal degrees of $\mathrm{tg}^2\theta$ lead to we get the matrix equation of the form

$$F^{(2m)} = \sum_{n=0}^{\infty} \frac{\tilde{A}_{2(n+m)+1}}{(2n+1)!}. \tag{8}$$

We denoted here

$$\tilde{A}_{2n+1} = \frac{A_{2n+1}}{(2R)^{2n+3}}(2n+1)!, \quad F^{(2m)} = (2m+1)! \tag{9}$$

The matrix of the equation (8) has the upper Jordan form [9]. The highest line of the matrix contains the elements equal to $1/(2n+1)!$, where $n = 0, 1, 2, \ldots$. The all subsequent non-zero elements follow from the first one by means of the shift on one step to the right so the main diagonal contain the only the units. Evidently the reverse matrix of (8) must have the Jordan form of the matrix also with the units in it's main diagonal. So the solution of the equation (8) may be found and written in the form

$$A_{2n+1} = \frac{(2R)^{2n+3}}{(2n+1)!}\sum_{m=0}^{\infty} M^{-1}_{nm} F^{(2m)}. \tag{10}$$

The reverse matrix in the expression (10) may be presented for one of the form

$$M^{-1}_{nm} = \frac{1}{2\pi i}\oint \frac{\varsigma^{2(n-m)}}{\mathrm{sh}\varsigma}\,d\varsigma. \tag{11}$$

Here the integration is in analytic region by arbitrary path, which surrounds all poles of under integral function [10]. This form of matrix is convenient for the analysis of the explicit expression of the matrix elements and for the proof of the statement (10).

Another method using the Fourier transformation more quickly leads to the solution of the equation (8). He is useful for the practical aims also. In this case for the explicit form of matrix elements we get the expression

$$\frac{A_{2n+1}}{(2R)^{2n+3}} = \frac{1}{2\pi}\int_{-\pi}^{\pi}\frac{\mu(x)}{sh(e^{-ix})}\frac{e^{-i(2n+1)x}}{(2n+1)!}dx \tag{12}$$

Here the function $\mu(x)$ is defined by the expression

$$\mu(x) = 1 + 2\sum_{m=1}^{\infty}(2m+1)!\cos 2mx. \tag{13}$$

Note the series presenting this function is diverging. Below because of this the function $\mu(x)$ is considered as an analogue of the generalized function or a functional having the sense only if he is integrated with another sufficiently smooth function. So all coefficients in (5) are defined. After substitution them in (5) and summing by $n$ we can get for the potential of the field $\varphi$

$$\varphi = -E_0 r\cos\theta\left[1 - \frac{(2R)^2}{\cos\theta\, r^2}\frac{1}{2\pi^2}\int_0^{\pi}du\int_{-\pi}^{\pi}dv\frac{\mu(v)}{sh(e^{-iv})}sh\left[\frac{2R}{r}(\cos\theta + i\sin\theta\cos u)e^{-iv}\right]\right] \tag{14}$$

The integral on variable $u$ here may be simplified because of the formula $sh(x+y) = sh(x)ch(y) + ch(x)sh(y)$ takes place. Indeed the resolution of the hyperbolic sinuous $sh(y)$ in the integral (14) in the series by degrees of $\cos(u)$ contains only the odd degrees which are vanishes. Rest integral of $u$ is expressing trough the Bessell's function of null order [8]. Then we get the expression

$$\varphi = -E_0 r\cos\theta\left[1 - \left(\frac{2R}{r}\right)^3\frac{1}{2\pi}\int_{-\pi}^{\pi}dv\mu(v)J_0\left(a\cdot e^{-iv}\right)\frac{sh(b\cdot e^{-iv})}{b\cdot sh(e^{-iv})}\right] \tag{15}$$

Here the parameters $a$ and $b$ are defined by the expressions

$$a = \frac{2R}{r}\sin\theta, \quad b = \frac{2R}{r}\cos\theta. \qquad (16)$$

On the sphere where $r = 2R\cos\theta$ the potential vanishes because of the identity

$$\frac{1}{2\pi}\int_{-\pi}^{\pi} dv\,\mu(v)J_0\left(\text{tg}\,\theta\cdot e^{-iv}\right) = \cos^3\theta. \qquad (17)$$

This statement may be proofed from the comparison of the resolutions left and right sides of (17) in degrees of $\text{tg}^2\theta$ taking into account the formula $\cos\theta = 1/\sqrt{1+\text{tg}^2\theta}$. The integral (15) in compact and explicit sight is not calculated. In a computer programs the function $\mu(x)$ is not convenient also. In this reason one define the function

$$h(x) = J_0(a\cdot x)\frac{\text{sh}(b\cdot x)}{b\cdot \text{sh}\,x}. \qquad (18)$$

Then we write it in form of the resolution through the Bessell function of null order

$$h(x) = \int_0^\infty a(k')J_0(k'x)\,dk'. \qquad (19)$$

The resolution coefficients may be found if we take into account the orthogonal conditions in [11] written as

$$\alpha\int_0^\infty J_m(\alpha\xi)J_m(\beta\xi)\xi\,d\xi = \delta(\alpha-\beta). \qquad (20)$$

Here $\delta(x)$ is Dirac's $\delta$- function, $m$ – the natural number. Then one may multiply both sides of (19) on $kxJ_0(kx)$ and integrate by variable of $x$. With the aid of (20) we get

$$a(k) = \int_0^\infty h(x)kxJ_0(kx)\,dx. \qquad (21)$$

The resolution of (19) is complete. To proof this statement it is necessary substitute the expression (21) in the (19) one taking into account the orthogonal condition (20) again. After substitution of the resolution (19) into the (15) then we take into account the boundary condition (17) and the definitions (16). As a result we can get the optimal expression for the potential wanted

$$\varphi = -E_0\, r\cos\theta\left[1-\left(\frac{2R}{r}\right)^3 \int_0^\infty xe^{-x} J_0(ax)\frac{\operatorname{sh}(bx)}{b\operatorname{sh}(x)}dx\right] \quad (22)$$

Note that the potential (22) obey the boundary conditions above stated. This follows also from the comparison of the identical formulas (22) and (15) in principle.

## 3. The distribution of the field potential, the surface density of the charge and the force acting on the sphere

For the visual representation of a counter plot of the potential (22) set up the scale of measurement of a distance to be equal diameter of the sphere. Formally this condition corresponds to the radius of the sphere $R = 0.5$ chosen in arbitrary linear units. At the Fig. 1 it is shown the counter plot in any plane passing trough the center of the sphere being normal to the initial plain of the problem.

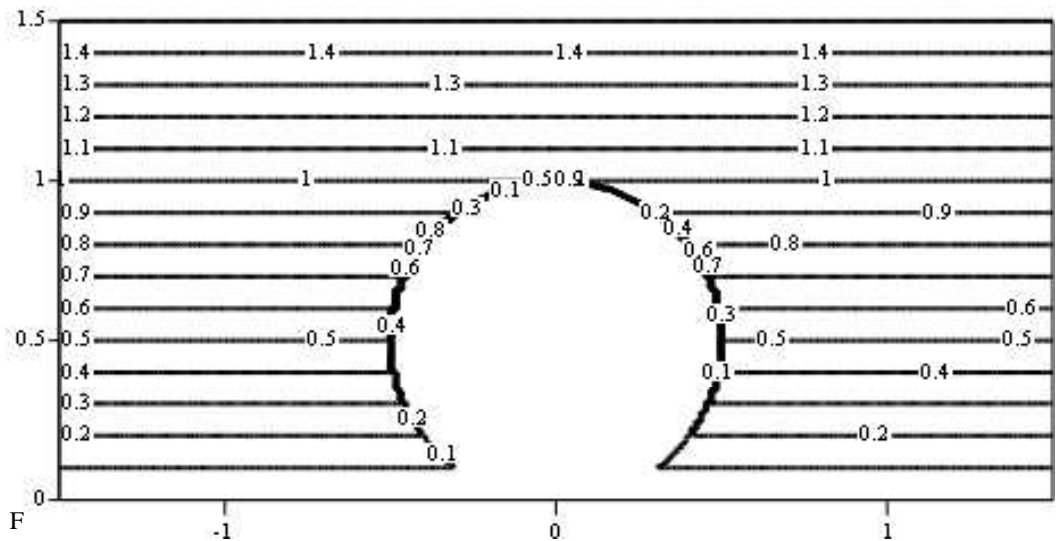

Figure 1: The counter plot of the potential with 200x200 points of divisions

The image here is constructed with the aid of MathCad (2000) Professional default computer program. We take in the vertical direction the region from 0 to 1.5 of diameters and the region from −1.5 to 1.5 of diameters in the horizontal direction. The potential lines are given in $-E_0\cdot 2R$ units and are separated in a relative value of 0.1. The lines are flowing in an expected manner. So far from the sphere they are parallel. Nearby the sphere they are strongly condensate in a small region. This one is seen only from the numerical signs. But they are seen as separately lines in the low part of Fig.2 where we have chosen 40x40 points of divisions.

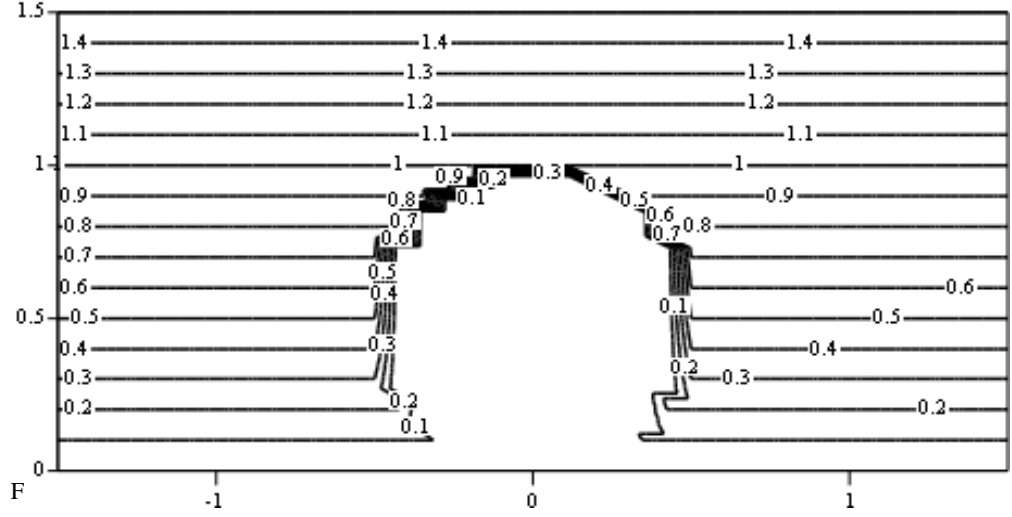

Figure 2: The counter plot of the potential with 40x40 points of divisions

At any point the electrical field components are found from the general gradient connection between field **E** and potential $\varphi$: $\mathbf{E} = -\,\mathrm{grad}\,\varphi$. So from (22) we may right now the radial $E_r$ and the angle $E_\theta$ components respectively $E_r = -\,\partial\varphi/\partial r$, $E_\theta = -\,(1/r)\,\partial\varphi/\partial\theta$. The component $E_z$ normal to the plane is $E_z = E_r \cos\theta - E_\theta \sin\theta$. At the plane points, where $\theta = \pi/2$, we have $E_z = -\,E_\theta$, because of condition $E_r = 0$. So in quadrature form the electrical field $E_z$ is

$$E_z = E_0 \left[ 1 - \left(\frac{2R}{r}\right)^3 \cdot \int_0^\infty \frac{u^2}{\mathrm{sh}(u)} e^{-u} J_0\left(\frac{2R}{r} u\right) du \right] \qquad (23)$$

Here $r$ is the distance from the contact point to the arbitrary point of the plane. The distribution of the surface charge density on the plane $\sigma_z$ we found from the expression $\sigma_z = E_z/4\pi$ (or from the expression $\sigma_z = \varepsilon_0 E_z$ in SI units, where the constant $\varepsilon_0 = 8.858\times10^{-12}$ ($C^2/N \cdot m^2$) is known as the permittivity of free space). At large distances from the sphere $r \gg 2R$ the electrical field is $E_z \approx E_0$. Here the surface charge density is uniform with value of density $\sigma_0$. Close to the sphere the surface charge density $E_z$ and the charge density de-crease (Fig. 3). In close proximity to the sphere $r \ll 2R$ the surface charge density and the field $E_z$ vanish according the law $E_z \approx E_0 (35/16)(r/2R)^6$ (Fig. 4).

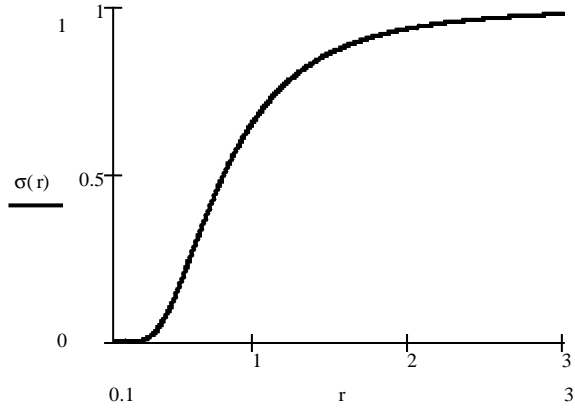

Figure 3: The dependence of the surface charge density $\sigma(r)$ on the plane in $\sigma_0$ units from the distance $r$ in the diameter 2R units

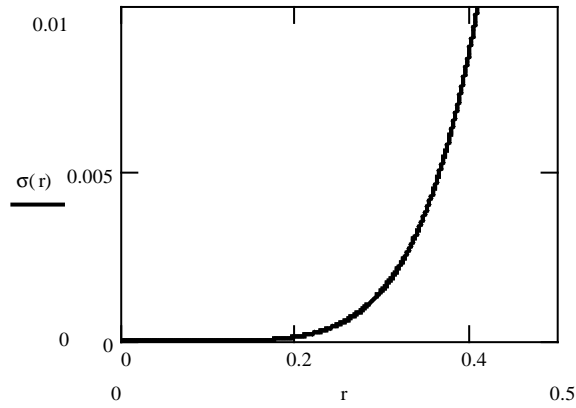

Figure 4: The dependence of the surface charge density $\sigma(r)$ on the plane in $\sigma_0$ units from the distance $r$ in the diameter 2R units in close proximity to the contact point

The normal $E_\perp$ and parallel $E_\parallel$ components to the sphere may be found from the expressions $E_\perp = E_r \cos\theta + E_\theta \sin\theta$, $E_\parallel = - E_r \sin\theta + E_\theta \cos\theta$. On the surface of the conducting sphere the condition $E_\parallel = 0$ takes place. Then the angle component $E_\theta$ is expressed in terms of the radial component $E_r$. In quadrature form for $E_r$ we may get the expression

$$E_r = E_0 \cos\theta \left[ -1 + 3\cos^2\theta + \frac{1}{\cos^3\theta} \int_0^\infty u^2 e^{-u} \operatorname{cth}(u) J_0(\operatorname{tg}\theta \cdot u)\, du \right] \qquad (24)$$

On the surface of the sphere the component $E_\perp$ is expressed through the $E_r$ as follows $E_\perp = E_r/\cos\theta$. So the surface charge density on the sphere $\sigma_s$ is $\sigma_s = E_\perp/4\pi$, (in SI we have $\sigma_s = \varepsilon_0 E_\perp$). At small angles $\theta \ll 1$ the surface charge density on the sphere is approximately four times more then $\sigma_0$. If the angle $\theta$ is rises the surface charge density on the sphere $\sigma_s$ decreases (Fig. 5).

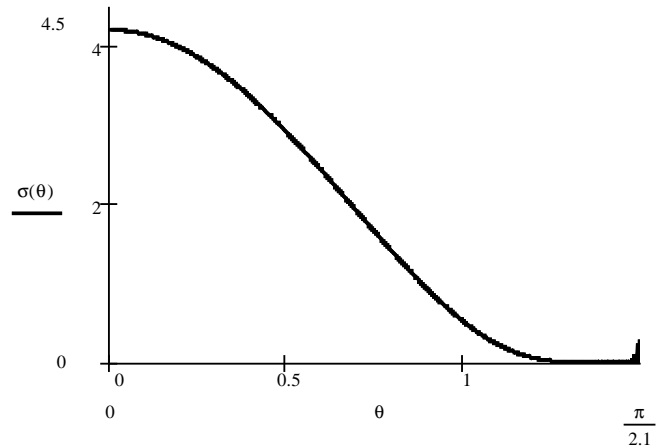

Figure 5: The dependence of the surface charge density $\sigma(\theta)$ on the sphere in $\sigma_0$ units from the place angle $\theta$

In angle region $\theta$ nearby to $\pi/2$ the surface charge density $\sigma(\theta)$ decreases according to law $378\cos^8\theta$ (Fig. 6). So from the region under the sphere the charge is pushed up and to the periphery.

The full charge $q$ on the sphere we may find by means of the integration the density $\sigma_s$ over the surface of the sphere. We get

$$q = \int_0^{\pi/2} \sigma_s 8\pi R^2 \sin\theta \cos\theta \, d\theta = 1.644 \, q_0. \tag{25}$$

Here we denoted the value of $q_0 = \sigma_0 4\pi R^2$. The sense is the charge of the sphere if her charge density were $\sigma_0$.

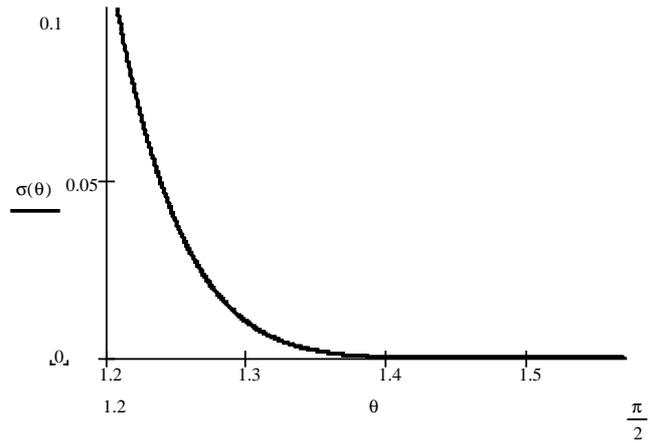

Figure 6: The dependence of the surface charge density $\sigma(\theta)$ on the sphere in $\sigma_0$ units from the place angle $\theta$ in close proximity to the contact point

According to the general rule [1] the force acting on the sphere may be found from the expression for radial surface density of the force equals to $(1/2)\sigma_s E_\perp$. From the symmetry of the task it follows that the force is directed along the z-axis. Expressed in SI in terms of the charge density $\sigma_s$ then after computing the integral - in terms of $q_0$ and $E_0$, the force is presented as

$$F = \frac{\pi R^2}{\varepsilon_0} \int_0^{\pi/2} \sigma_s^2 \sin 4\theta \, d\theta = 1.368 \, q_0 E_0 \qquad (26)$$

To estimate the addition of the jumping mechanisms to the transfer of charge let us consider two plane electrodes with the vertical orientation of the planes choose so to neglect the force of the gravity. Suppose that among the planes is the ideal gas of the conducting nanoparticles like the molecules of $C_{60}$. If the distance $d$ between the electrodes is more than the diameter $2R$ the nanoparticle contacting with an electrode will receive the charge (25). If the contact failed the nanoparticle will be acted the force equal in order of the value $qE_0$. This force causes the transfer the charge of nanoparticle $C_{60}$ with mass $m = \mu/N_A$, where $\mu$ and $N_A$ are respectively the molecular mass and Avogadro's number, from one plate to another in time $\tau$ which is estimated by formulae

$$\tau = \sqrt{\frac{2dm}{1.644 \cdot 4\pi R^2 \cdot \varepsilon_0 \cdot E_0^2}}. \tag{27}$$

If $d = 100$nm, $2R = 1$nm и $E_0 = U/d = 10^9$ V/m, where U = 100 V is bias voltage we find for the frequency $\nu = 1/2\tau = 6.9 \cdot 10^9$ Hz. This value of the frequency is sufficiently high to be taking into account. Note that this estimation did not resign the concrete mechanisms of the shutlle transfer considered in [6].

If the force of gravity is taking into account the condition of loss contact with the plane is defined from the equality of the force (26) to the weight of the sphere. In this case the estimation of the minimal value of the electric field may be found from the formula

$$E_0 = \sqrt{\frac{\rho g R}{3 \cdot 1.368 \varepsilon_0}}. \tag{28}$$

Here $\rho$ and g are the density of mass and acceleration of gravity respectively. So for the steel material of the sphere with R = 1mm we get the value $E_0 \geq 14.51$ (кV/cm). If we put the R = $1\mu$m one finds that $E_0$ is obeying the condition $E_0 \geq 459$ (V/cm). The preliminary experiments with the carbon and the graphite powder showed that here in our case it is possible an variant of the well know Millikan's oil-drop experiment described in textbooks. The difference is the particle is observed not at the moment of being suspended in the ear when the forces balanced but at the moment when it is levitating in the moment of the loss of their contact with conducting plane. For the nanoparticles this experiment may be realized in the fields less than $E_0 \approx 14.51$ (V/cm). He is of great moment for the measurements a sizes and a mass in the field $E_0$ known. It may be this experiment be realized with the conducting structures of graphene [12] and $C_{60}$ molecules. Even more this task may by useful for solving some important problems of Fullern's molecules lying in the graphene plane [13]. It will be noted that if the force (26) known we may then separate the micro- and nanoparticles in principle.